\documentclass[pre,twocolumn]{revtex4}
\usepackage{psfrag}
\usepackage{amssymb,amsmath,amsthm}
\usepackage[dvips]{graphicx}
\usepackage{verbatim}
\usepackage[colorlinks=true,linkcolor=blue,citecolor=blue]{hyperref}
\usepackage[caption=false]{subfig}

\begin{document}

\title{Synchronization of coupled noisy oscillators: Coarse-graining from continuous to discrete phases}
\author{Daniel Escaff$^1$, Alexandre Rosas$^2$, Ra\'ul Toral$^3$, and Katja Lindenberg$^4$}
\affiliation{
$^1$Complex Systems Group, Facultad de Ingenier\'{i}a y Ciencias
Aplicadas, Universidad de los Andes, Avenida Monse\~{n}or \'{A}lvaro del Portillo N$^{\text{o}}$ 12.455, Las Condes,
Santiago, Chile. \\$^2$Departamento de F\'{i}sica, CCEN, Universidade Federal da Para\'{i}ba,
Caixa Postal 5008, 58059-900, Jo\~{a}o Pessoa, Brazil \\
$^3$IFISC (Instituto de F\'isica Interdisciplinaria y Sistemas Complejos), CSIC-UIB, E-07122 Palma de
Mallorca, Spain\\
$^4$Department of Chemistry and Biochemistry and BioCircuits Institute,
University of California San Diego, La Jolla, California 92093-0340, USA}

\begin{abstract}
The theoretical description of synchronization phenomena often relies on coupled units of continuous time noisy Markov chains with a small number of states in each unit. 
It is frequently assumed, either explicitly or implicitly, that coupled discrete-state noisy Markov units can be used to model mathematically more complex coupled noisy continuous phase oscillators. In this work we explore conditions that justify this assumption by coarse-graining continuous phase units. In particular, we determine the minimum number of states necessary to justify this correspondence for Kuramoto-like oscillators. 
\end{abstract}


\maketitle

\section{Introduction}
\label{sec:introduction}

Synchronization phenomena have been intensely studied for decades, in part because of the roles such phenomena play in chemical systems, laser arrays, cellular biology models, and neural networks to name just a few 
(see Refs.~\cite{Strogatz,Pikovsky,Manrubia,Acebron} for extensive reviews). One of the most extensively studied models is that proposed by Kuramoto in 1975~\cite{Kuramoto1}, a model that has become paradigmatic for the description of many synchronization phenomena.  Originally the model was applied to an interacting population of oscillators with randomly distributed frequencies. When the interaction is sufficiently strong, most of the units in the array synchronize their dynamics to a single frequency which may differ from the natural frequency of any one of the synchronized oscillators, and also to equal phases.  

Many variants of the original model have been introduced over the years to study different effects in different physical and biological systems, too many to list here (for an extensive review, see Ref.~\cite{Acebron}). We specifically mention the inclusion of fluctuations, because of their central role in our studies. 
Noise leads to disorder, so in the presence of noise the interactions that in its absence may be strong enough to lead to frequency or to phase synchronization must in general be stronger for synchronization to occur. In all of these models, the form and range of the interactions has varied greatly in the literature.
  
Beyond the Kuramoto model, many different models for synchronization have been proposed, ranging from arrays of continuous oscillatory and excitable units to discrete models. For instance, over the past decade coupled maps have attracted a great deal of attention~\cite{Pikovsky,Gade}.
Recently, arrays of coupled stochastic units each with a discrete set of states but, in contrast with maps, with continuous time have increased in popularity as a simpler paradigm for 
synchronization~\cite{Prager1,Prager2,Kouvaris1,Kouvaris2,Wood1,Wood2,Wood3,Wood4,Assis1,Assis2,Escaff1,Escaff2,Pinto,Escaff3,Rosas,Wood5}. Even though these discrete-state oscillator models may be motivated by discrete processes (for example, protein degradation~\cite{Lafuerza1,Lafuerza2,Lafuerza3}), it has been claimed that they can also be used to model a coarse-grained phase space of continuous noisy oscillators. For instance, 
Prager et al.~\cite{Prager2,Kouvaris1} established a link between a globally coupled ensemble of excitable units described by the FitzHugh-Nagumo equations with additive white noise, and a coupled array of 3-state non-Markovian stochastic oscillators.

Our own work has focused on arrays of 2-state and of 3-state stochastic oscillators. The transitions between the states of individual units are governed by a rate process. This rate process might be Markovian~\cite{Wood1,Wood2,Wood3,Wood4,Pinto,Escaff3,Rosas} or might involve distributed delays (such as, for instance, a refractory period)~\cite{Escaff1,Escaff2}. Interactions among units in our model appear as a dependence of the transition rates of a particular unit on the states of the other units to which it is coupled.

The goal of the work presented herein is to address the following two questions: (1) Under what conditions can we describe the dynamics of Kuramoto-like coupled noisy oscillators as periodic continuous-time Markov chains? In other words, when can we model continuous-phase stochastic dynamics as discrete-phase models in which the transitions between the discrete states are governed by memoryless rate processes? (2) Is there a lower limit to the discretization of the continous noisy oscillators? In other words, how many discrete states are necessary to capture the essential synchronization features of the continuous system? The popularity of three-state models leads us to explore whether the synchronization properties of coupled three-state Markovian units in any way capture those of the continuous oscillator system. To arrive at some answers to these questions, in Sec.~\ref{sec:review} we present the continuous phase model that is the starting point of our analysis. It is an array of Kuramato-like oscillators with additive noise and a generalized nonlinear interaction. We start with the full amplitude equations, but will always work in the limit where the phase equations alone provide a valid description of the important dynamics. 
For the sake of simplicity we consider a globally coupled ensemble of identical oscillators (all with the same natural frequency), and thus focus on the phase synchronization phenomenon. In Sec.~\ref{sec:coarsegraining} we perform the coarse-graining of the phase space and discuss the conditions under which the dynamics can be modeled as a periodic Markov chain. Here we also discuss the questions associated with the three-state systems.
Finally, in Sec.~\ref{sec:summary} we present our concluding remarks. We also include two appendices with technical details of our calculations.

\section{Model and brief review of phase synchronization}
\label{sec:review}

Our starting point is an ensemble of $N$ identical noisy oscillators described by the complex time-dependent dimensionless amplitudes $A_s(t)$, with $s\in \left\{1, ..., N\right\}$. These amplitudes are governed by the equations of motion
\begin{equation}
\dot{A}_s = J f_0 ( \left|A_s\right|^2)A_s + K f(\left|\mathcal{R}\right|^2)\mathcal{R} + \sqrt{\eta}\zeta_s(t).
\label{AmplitudeOsc}
\end{equation} 
The overdot indicates a derivative with respect to time. $J$ is a real positive parameter that governs the internal dynamics of each oscillator. For the function $f_0$, which describes this internal dynamics, we take the normal form of a supercritical Hopf bifurcation,
\begin{equation}
 f_0 ( \left|A_s\right|^2) = 1 - \left|A_s\right|^2.
\end{equation} 
For simplicity, we take all parameters to be real and have scaled out irrelevant constants. The oscillators are identical, and we have removed the natural frequency of oscillation of each unit, that is, we are working in a moving framework. In the usual language of the Kuramoto model, the frequency distribution of the oscillators is $g(\omega) = \delta(\omega)$, where the $\delta$-function is appropriate for the continuous variable $\omega$ (and below also for the continuous time $t$). Therefore, the internal dynamics of each oscillator tends to set $\left|A_s\right| = 1$, with an arbitrary phase. 

The second term on the right hand side of Eq.~(\ref{AmplitudeOsc}) accounts for the interaction between the oscillators. The coupling strength is quantified by $K>0$. The interaction is assumed to be global (all-to-all interaction), with the customary Kuramoto order parameter given by the average amplitude as a function of time,
\begin{equation}
\mathcal{R} = \frac{1}{N}\sum_{s^{\prime} = 1}^{N} A_{s^{\prime}}.
\label{order1}
\end{equation} 
The original Kuramoto model \cite{Kuramoto1} is recovered if we set $f(\mathcal{R})$ equal to $1$, so that the global interaction is given by $K\mathcal{R}$. The function $f$ accounts for a nonlinear interaction between the oscillators via $\mathcal{R}$. The advantage of including a general nonlinear function $f$ in the interaction will be clear when we subsequently perform the coarse-graining operations.

The third term on the right hand side of Eq. (\ref{AmplitudeOsc}) is a complex additive noise of intensity $\eta$. This term models the fluctuations. The noise is of the form
\begin{equation}
\zeta_s(t) = \zeta_R^{s}(t) + i\zeta_I^{s}(t),
\end{equation} 
where $ \zeta_R^{s}(t)$ and $ \zeta_I^{s}(t)$ are independent real Gaussian white noises of  zero mean and correlation functions
\begin{equation}
\left\langle \zeta_R^{s}\left(t\right)\zeta_R^{s^{\prime}}\left(t^{\prime}\right) \right\rangle = \left\langle \zeta_I^{s}\left(t\right)\zeta_I^{s^{\prime}}\left(t^{\prime}\right) \right\rangle =\delta_{ss^{\prime}}\delta\left(t-t^{\prime}\right), 
\label{zeta1}
\end{equation} 
\begin{equation}
\left\langle \zeta_R^{s}\left(t\right)\zeta_I^{s^{\prime}}\left(t^{\prime}\right) \right\rangle = 0.
\label{zeta2}
\end{equation} 
Here  $\delta_{ss'}$ is the Kronecker delta appropriate for the discrete variable $s$.
We note that the form of Eq.~(\ref{AmplitudeOsc}) respects the phase invariance, that is, the  
equation is invariant under the transformation 
\begin{equation}
\forall s\in \left\{1, ..., N\right\};~~   A_s \rightarrow A_s e^{i\phi_0},
\end{equation} 
with $\phi_0$ constant, but the equation is otherwise quite general.

\subsection{Phase equation}

We consider the parameter range $J\gg K$ and  $J\gg\eta$, so that the time scale of the internal dynamics of each oscillator dominates over (i.e. is shorter than) that of the interactions between the oscillators. Then, after a fast transient defined by the internal dynamics, we have that $\left|A_s\right|\approx 1$. After that, the phase of each oscillator varies as a function of time on a slower time scale defined by the interactions (albeit with very rapid fluctuations). On this longer time scale we can write $A_s\approx e^{i\phi_s}$. The dynamics specified by Eq.~(\ref{AmplitudeOsc}) can then be reduced to the phase equation~\cite{Kuramoto2}
\begin{equation}
\dot{\phi}_s = K F(r)\sin\left(\psi - \phi_s\right) + \sqrt{\eta}\xi_s\left(t\right).
\label{PhaseOsc}
\end{equation} 
Here we have defined a Kuramoto order parameter $R$ which follows directly from Eq.~(\ref{order1}), 
\begin{equation}
R = \frac{1}{N}\sum_{s = 1}^{N} e^{i\phi_{s}} \equiv r e^{i\psi},
\label{KOP}
\end{equation}
from which we extract the real phase variable $\psi$, and
\begin{equation}
F(r) = f (r^2)r,
\label{Gfunction}
\end{equation}
where $r\in[0,1]$ and $\phi_s\in[0,2\pi]$ are also real.  The noise $\xi_s(t) = -\sin\phi_s\zeta_R^{s}\left(t\right) + \cos\phi_s\zeta_I^{s}\left(t\right)$ is again Gaussian and white, with zero mean and correlation function that follows directly from Eqs.~(\ref{zeta1}) and (\ref{zeta2}),
\begin{equation}
\left\langle \xi_{s}\left(t\right)\xi_{s^{\prime}}\left(t^{\prime}\right) \right\rangle =\delta_{ss^{\prime}}\delta\left(t-t^{\prime}\right).
\end{equation} 

\subsection{Mean field nonlinear Fokker-Planck equation}

The order parameter $R$ can be written as
\begin{equation}
R = \int_{0}^{2\pi} n\left(\phi, t\right)e^{i\phi} d \phi,
\end{equation} 
where we have introduced the density of oscillators with phase $\phi$,
\begin{equation}
 n\left(\phi, t\right) =  \frac{1}{N}\sum_{s = 1}^{N} \delta\left[\phi - \phi_{s}\left(t\right)\right].
\end{equation} 
In the thermodynamic limit $N\rightarrow\infty$, 
\begin{equation}
\lim_{N\rightarrow\infty} n\left(\phi, t\right) =  \rho\left(\phi, t\right),\label{MFapprox}
\end{equation} 
where $\rho\left(\phi, t\right)d\phi$ is the probability that the phase of an oscillator lies in the interval $[\phi, \phi + d\phi]$ at time $t$.

In the thermodynamic limit, the stochastic phase equation (\ref{PhaseOsc}) (which is an equation of the Langevin form) can be replaced by a nonlinear Fokker-Planck equation (see Ref.~\cite{Acebron} for a detailed derivation using the path integral formalism),
\begin{equation}
\frac{\partial\rho}{\partial t} = \frac{\eta}{2}\frac{\partial^2\rho}{\partial\phi^2} - K\frac{\partial}{\partial\phi}\left\{\rho\Omega\left[\rho,\phi\right]\right\} ,
\label{NLFPE}
\end{equation} 
where the second derivative term on the right is the diffusion term, and where the drift contains
\begin{equation}
\Omega\left[\rho,\phi\right] =  F(r\left[\rho\right])\sin\left(\psi\left[\rho\right] - \phi\right) ,
\label{5b}
\end{equation} 
with
\begin{equation}
R=r\left[\rho\right]e^{i\psi\left[\rho\right]} \equiv  \int_{0}^{2\pi} \rho\left(\phi, t\right)e^{i\phi} d \phi.
\end{equation} 

\subsection{Phase synchronization}

Since all of our oscillators have the same frequency ($\omega = 0$ in the moving frame), synchronization in this framework corresponds to the tendency of the oscillators to have the same phase. The desynchronized state corresponds to a uniform distribution of phases,
\begin{equation}
\rho(\phi) = \frac{1}{2\pi}.
\label{eq:desync}
\end{equation} 
This choice obeys the normalization condition 
\begin{equation}
\int_{0}^{2\pi}\rho(\phi) d\phi=1.
\label{normcont}
\end{equation}
The dynamics described by the phase equation (\ref{PhaseOsc}) contains two competing trends: the fluctuations, which tend to desynchronize the system and stabilize the state described in Eq.~(\ref{eq:desync}), and the attractive interactions that tend to synchronize the system. When the coupling strength is weaker than a critical value ($K<K_c$), the desynchronized state is stable, while an interaction stronger than this critical value ($K>K_c$)
destabilizes the desynchronized state, leading to a phase transition to synchronization. Calculations of the critical point have been widely documented in the literature~\cite{Acebron,Kuramoto2}, and we see no point in repeating them here. Instead, here we consider the continuous limit of the coarse-graining procedure presented in  Sec.~\ref{sec:coarsegraining} and detailed in Appendix A. This limit is in turn detailed in Appendix B, where we show that this critical value is given by
\begin{equation}
K_c = \frac{\eta}{f(0)}
\label{CriticalK-1}.
\end{equation}
As would be expected, stronger fluctuations require a stronger interaction for synchronzation to occur, that is, the critical value of the coupling parameter increases with increasing noise intensity $\eta$.

In the thermodynamic limit the absence of synchronization is associated with a vanishing order parameter, $R = 0$ . Away from the thermodynamic limit $R$ fluctuates to values away from zero because of the finite number of oscillators, but remains small as long as $N$ is not exceedingly small. The occurrence of self-organization is characterized by a non-vanishing order parameter. Near the critical point (\ref{CriticalK-1}), the mean-field evolution of the order parameter can be approached via the normal form (see Refs.\cite{Acebron,Kuramoto2,Bonilla} and Appendix B)
\begin{equation}
\dot{R} = \left(\alpha -\beta\left|R\right|^2\right)R\label{NormalForm-1},
\end{equation}
with
\begin{align}
\alpha =& \frac{f(0)}{2}\left(K - K_c\right), 
\nonumber\\
\beta =& \frac{K_c}{2} \left(\frac{1}{2}f(0) - f^{\prime}(0)\right),
\label{alphabeta}
\end{align}
where $f^{\prime}(0) \equiv \left.(df(r)/dr)\right|_{r=0}$. For the original Kuramoto model \cite{Kuramoto1}, $f=1$, the transition to synchronization is always supercritical because $\beta>0$. The presence of a nonlinear interaction $f(r)$ might induce a change in sign of $\beta$ and might therefore lead to a subcritical transition. The function $f(r)$ affects the critical coupling strength in Eq.~(\ref{CriticalK-1}) simply as a renormalization of this critical value. The main role of $f(r)$ is to increase or decrease the dispersion of the phases around a common value in the synchronized regime. 
The qualitative effect of including the function $f$ can be seen in the interaction term in the phase equation (\ref{PhaseOsc}). In the synchronized phase a function $f(r)$ that increases with increasing $r$ increases the attractive interactions among the oscillators, leading to a decrease of the dispersion of phases around a common value. In contrast, if $f(r)$ decreases with increasing $r$, this tendency will be weakened and, since fluctuations lead to a greater dispersion of the phases, the synchronization will be coarser. A decreasing function might thus be a better candidate for a successful increasingly coarse coarse-graining procedure.

\subsection{Numerical analysis}

\begin{figure}
\includegraphics[width =2.8in]{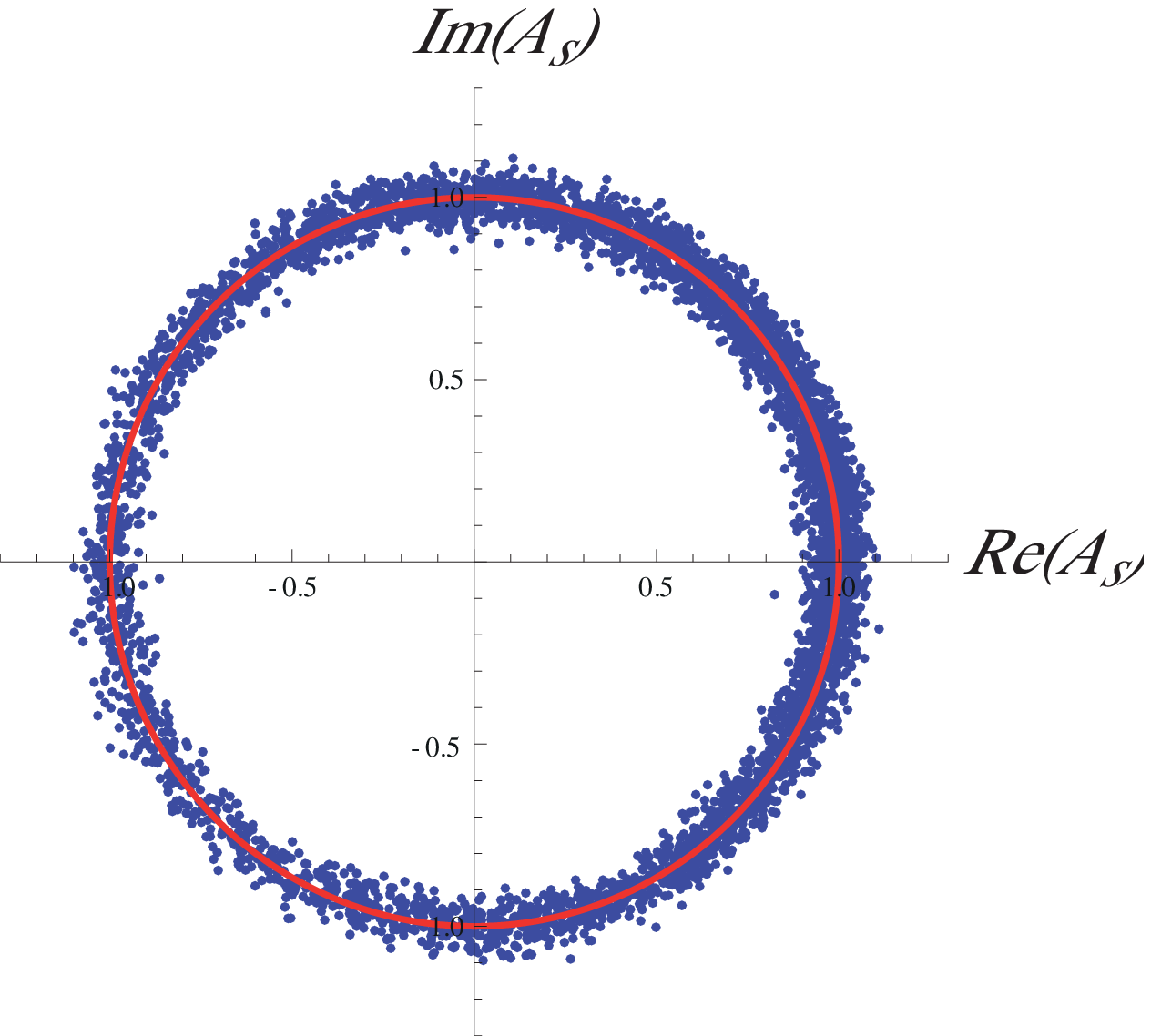}\\
\includegraphics[width =2.8in]{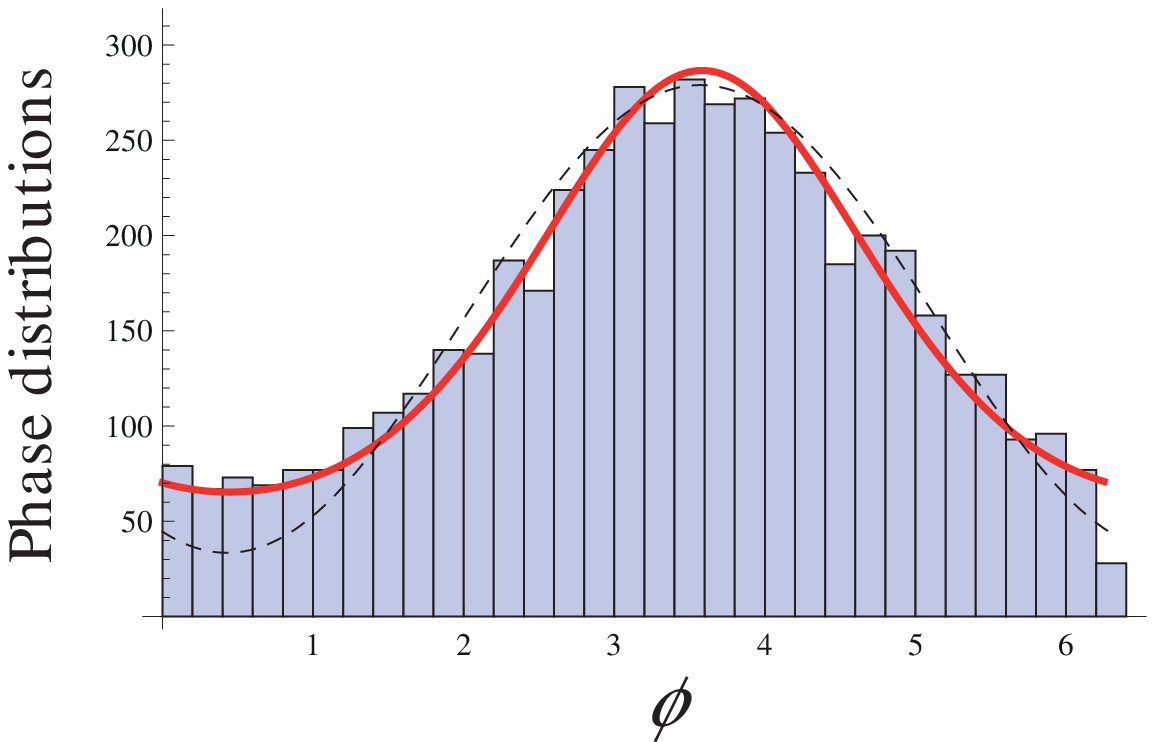}
\caption{Numerical study using the model function (\ref{fModel-1}) for $f(r)$, with $a=0.3$. The noise intensity is $\eta=0.98696$, while the coupling strength is $K=1.5708 > K_c=0.98696$. This places us well into the synchronized regime.
Upper panel: Direct numerical simulation of Eq.~(\ref{AmplitudeOsc}), with $J=200$ and $N = 5000$. The  points represent the complex amplitude of each oscillator in the complex plane at the instant $t=50$, having started at time $t=0$ with all amplitudes at $A_s=0$. The red curve is the circle of unit radius. Bottom panel: Histogram of the phases of the oscillators. The thick continuous (red) curve shows a numerical computation of the steady state of the nonlinear Fokker-Planck equation (\ref{NLFPE}). The thin(black) dashed curve shows the analytic estimation (\ref{NFapprox}). All the distributions are normalized to the number of oscillators, that is, multiplied by the factor $2N\pi/B$, where $B=32$ is the number of bars of the histogram. The width of the histogram, also apparent in the width of the ring in the upper panel, is due to the fluctuations and to the rapid decrease of the coupling function. }
\label{fig01}
\end{figure}

For numerical simulations we must choose a particular form for the function $f(r)$. As stated above, our goal of coarse-graining the phase space is better achieved if we choose a decreasing function of $r$. We have tested a number of different functions and have determined that the particular form and parameters in it 
are not as important as is simply the choice of almost any decreasing function of $r$. We use the exponentially decreasing function 
\begin{equation}
f(r) = \exp\left(-\frac{r}{a}\right) \label{fModel-1},
\end{equation}
with $a$ a positive parameter. Note that when $a\rightarrow\infty$,  $f\rightarrow 1$ and we recover the standard Kuramoto model.

In the upper panel of Fig.\ref{fig01} we show, after a transient, the state of an ensemble of $5000$ interacting oscillators governed by Eq.~(\ref{AmplitudeOsc}). The points represent the complex amplitude of each oscillator in the complex plane. As expected for large $J$ ($J=200$ in all our simulations), after a transient the amplitudes settle around a circle of unit radius (red line). We have chosen a coupling strength $K\gg K_c$ (see caption) so that the array of oscillators is in the synchronized regime. The system exhibits coarse phase clustering, that is, the synchronization is not perfect. This is attributed to the fluctuations  that tend to disperse the phases, and to the rapid decrease of the function (\ref{fModel-1}).

The bottom panel of Fig.\ref{fig01} displays the structure of a phase cluster. The figure shows the histogram of the phases of the oscillators in a cluster obtained from a numerical simulation of Eq.~(\ref{AmplitudeOsc}). The red continuous curve is a numerical computation of the steady state of the nonlinear Fokker-Planck equation (\ref{NLFPE}) . This steady state fits the simulation results very well even though the nonlinear Fokker-Planck equation relies on two approximations: the reduction of the amplitude and phase equation Eq.~(\ref{AmplitudeOsc}) to Eq.~(\ref{PhaseOsc}) for only the phases, and the mean field assumption that $N\to\infty$ although we have a finite number of oscillators. The dashed curve shows an analytic estimation of the steady state solution of the nonlinear Fokker-Planck equation, obtained from the normal form approach (see Appendix B):
\begin{equation}
\rho_{st}\left(\phi\right) \approx \frac{1}{2\pi}\left(1 +2\sqrt{\frac{2a(K-\eta)}{\eta(a+2)}} \cos(\phi + \psi)\right).\label{NFapprox}
\end{equation} 
Here the phase $\psi$ is a constant that defines the phase-cluster location in the unit circle and is fixed by the initial condition and the particular realization of the fluctuations. For finite $N$, the location of this phase in the unit circle is random.  This random phase is not captured by the steady state solution of Eq.~(\ref{NLFPE}) because in the mean field approximation (\ref{MFapprox}) the limit $N\rightarrow\infty$ is applied before the limit $t\rightarrow\infty$. In general these limits do not conmute, and finite size effects are lost in the mean field equation Eq.~(\ref{NLFPE}). In any case, sufficiently large $N$ allows for the formation of a phase-cluster which is associated with a non-vanishing value of the order parameter $R$. This fact is independent of the order of the limits.

\section{Coarse-graining phase space}
\label{sec:coarsegraining}

Instead of the continuous phase $\phi$, we move on to a discrete set of $M$ groups of phases discretely and equidistantly centered around the circle of unit radius seen in the top panel of  Fig.~\ref{fig01}.  That is, 
\begin{equation*}
\phi\in \left[0,2\pi\right]~ \rightarrow ~ \phi \in \left\{j\Delta\phi\right\}_{j=0}^{M-1},
\end{equation*} 
where
\begin{equation}
\Delta\phi = \frac{2\pi}{M}.
\label{Dphi}
\end{equation} 
We then convert the nonlinear continuous Fokker-Planck equation (\ref{NLFPE}) to the finite difference equation
\begin{align}
\dot{P}_j =& \frac{\eta}{2(\Delta\phi)^2}\left(P_{j+1} + P_{j-1} -2P_j\right)
\nonumber\\
& - \frac{K}{2\Delta\phi}\left(\Omega_{j+1}P_{j+1} - \Omega_{j-1}P_{j-1}\right) ,
\label{NLME-1}
\end{align}
with periodic boundary conditions $(M-1) + 1 \rightarrow 0$ and $0 - 1 \rightarrow (M-1)$.
Here $P_j (t)$ represents the probability of finding a unit in the phase group $j\Delta\phi$ (henceforth called state $j$), at time $t$, and  
\begin{equation}
\Omega_j = F(r)\sin\left(\psi - j\Delta\phi\right),
\label{bound}
\end{equation} 
with
\begin{equation}
R = re^{i\psi} = \sum_{j=0}^{M-1}P_j e^{i j \Delta\phi}.
\label{OPmf}
\end{equation} 
This is the associated version of the mean field order parameter. Note that in our new discrete state space the normalization of the probability is [cf. Eq.~(\ref{normcont})]
\begin{equation*}
\sum_{j=0}^{M-1}P_j = 1.
\end{equation*} 
The desynchronized state now is
\begin{equation}
P_j = \frac{1}{M}
\label{Uniform2}
\end{equation} 
for all $j$.

\subsection{Periodic continuous-time Markov chain model}

Equation (\ref{NLME-1}) can be written in the form
\begin{align}
\dot{P}_j =& - \left(w_{j\rightarrow j+1} + w_{j\rightarrow j-1}\right)P_j
\nonumber\\
& + w_{j +1 \rightarrow j}P_{j+1}  + w_{j - 1 \rightarrow j}P_{j-1},
\label{NLME-2}
\end{align}
where we define the transition rates
\begin{equation}
w_{j\rightarrow j\pm 1} =  \frac{\eta}{2(\Delta\phi)^2} \mp \frac{K}{2\Delta\phi}\Omega_{j}.
\label{Rate}
\end{equation} 
Equation (\ref{NLME-2}) is a master equation for a discrete state system, where the transitions between the $M$ states are governed by the rates (\ref{Rate}). Only nearest-neighbor transitions ($j\rightarrow j\pm 1$) are allowed. We stress that Eq.~(\ref{NLME-2}) is nonlinear because the rates (\ref{Rate}) depend on the $P_j$ via $\Omega_j$ [see Eqs.~(\ref{bound}) and(\ref{OPmf})].

We have arrived at the following point in this development: the stochastic dynamics of a single Kuramoto oscillator described by  the continuum equation Eq.~(\ref{AmplitudeOsc}) may be converted to the problem of a continuous-time Markov chain of  $M$ states, with periodic boundary conditions. These states model the internal structure of each stochastic oscillator. 
The interaction between the oscillators is contained in the dependence of the rates (\ref{Rate}) on the global state of the system. An ensemble of $N$ of these $M$-state oscillators is characterized by the densities
\begin{equation*}
n_j\left(t\right) = \frac{N_j\left(t\right)}{N},
\end{equation*} 
where $N_j\left(t\right)$ is the number of oscillators in state $j \in \left\{0, ..., M-1\right\}$ at time $t$. The rates (\ref{Rate}) depend on the global state $\left\{n_j\left(t\right)\right\}_{j=0}^{M-1}$
via the $\Omega_j$, which depend on the finite size order parameter 
\begin{equation}
R = re^{i\psi} = \frac{1}{N}\sum_{j=0}^{M-1}N_j e^{i j \Delta\phi} = \sum_{j=0}^{M-1}n_j e^{i j \Delta\phi}.\label{OPn}
\end{equation} 

In the thermodynamic limit,
\begin{equation*}
\lim_{N\rightarrow\infty} n_j\left(t\right) = \left\langle n_j\left(t\right)\right\rangle= P_j\left(t\right),
\end{equation*} 
and the order parameter (\ref{OPn}) converges to the mean-field version (\ref{OPmf}). In this limit, the infinite size ensemble of $M$-state units is described by the nonlinear master equation (\ref{NLME-2}), in the same spirit as the nonlinear Fokker-Planck equation (\ref{NLFPE}) description of the continuum units. If $N$ is finite,  we need to work with a set of coupled Langevin equations, as in Ref. \cite{Pinto,Rosas}. However, except for our numerical simulations, finite size effects are beyond the scope of this paper; here we focus on the mean-field theory. We note that  first taking the limit $N\rightarrow\infty$, and then the limit $t\to\infty$ to study the steady states of (\ref{NLME-2}) in general does not commute with taking these two limits in the opposite order \cite{Pinto}.

The continuous-time Markov-chain modeling scheme rests on the assumption that the rates (\ref{Rate}) are positive, which is not trivially satisfied. 
This requirement imposes constraints: $w_{j\rightarrow j\pm 1}$ is positive for any initial condition and realization of the fluctuations only if
\begin{equation}
K\Omega_{max} < \frac{\eta}{\Delta\phi},
\label{kmax}
\end{equation} 
where $\Omega_{max}$ is the maximum value of $\left|\Omega_{j}\right|$ for any $j$ in any realization of the evolution of the system. An evident bound is obtained from Eq.~(\ref{bound}) by noting that the maximum value of the sine is unity,
\begin{equation}
\Omega_{max} < F_{max} = \max_{r\in[0,1]} \left\{F(r)\right\}.
\label{omegamax}
\end{equation} 
Fixing the noise intensity and using the coupling strength $K$ as the control parameter, we impose the condition 
%Hence, fixing the noise intensity and using the coupling strength $K$ as the control parameter, the combination of Eqs.~(\ref{kmax}) and (\ref{omegamax}) limits the coupling as follows:
\begin{equation}
K < K_{max} = \frac{\eta}{F_{max}\Delta\phi}.\label{Kmax}
\end{equation} 
Equations~(\ref{Kmax}) and (\ref{omegamax}) ensure that condition (\ref{kmax}) is satisfied.
In the limit $\Delta\phi\rightarrow 0$ (continuous phase space), there is no upper limit on the coupling ($K_{max}\rightarrow\infty$), as expected from our earlier analysis. 

\subsection{Phase synchronization in the continuous-time Markov chain model}

The desynchronized state (\ref{Uniform2}) becomes unstable when the coupling strength $K$ exceeds the critical value (see Appendix A)
\begin{equation}
K_c = \frac{\eta}{f(0)}\left(\frac{\tan\left(\Delta\phi/2\right)}{\Delta\phi/2}\right)
\label{CriticalK-2}.
\end{equation}
This converges to the value (\ref{CriticalK-1}) when $\Delta\phi\rightarrow 0$ (continuous state space limit). 
At the opposite extreme, when $\Delta \phi = \pi$ (which corresponds to only two states on the circle, ($M=2$), Eq.~(\ref{CriticalK-2}) gives $K_c=\infty$, so that no instability to a synchronized state is observed. To observe synchronization within the framework of the continuous-time Markov chain picture, we must impose the condition
\begin{equation*}
K_c < K_{max},
\end{equation*}
which implies [see Eqs.~(\ref{Dphi}) and (\ref{Kmax})]
\begin{equation}
M > \frac{\pi}{\arctan\left(\frac{f(0)}{2F_{max}} \right)}.
\label{minM}
\end{equation}
For the standard Kuramoto model, $f(r)=1$ and $F(r)=r$, that is,  $f(0)/F_{max} = 1$. Coarse-graining for this model requires that $M>\pi/\arctan(1/2)$, that is, $M\geq 7$. Discretization with smaller $M$ within the Kuramoto model can not be interpreted as a Markov chain (the resulting ``transition rates" may not be positive). 

Coarser graining can be obtained by moving away from the Kuramoto model and considering ratios $f(0)/F_{max}<1$. For the model function (\ref{fModel-1}), maximizing the coupling $F(r)$ [cf. Eq.~(\ref{Gfunction})] yields
\begin{equation*}
\frac{f(0)}{F_{max}} = 
\left\{
  \begin{array}{cc}
 e^{1/2}\sqrt{2/a}~~~~  &\text{if}~~ a<2 \\
 e^{1/a}~~~~  &\text{if}~~ a>2  \\
  \end{array}
\right. .
\end{equation*}
This allows us to arbitrarily decrease the number of states in each unit. The limiting case $M=3$ can be reached by choosing $a<e/6\cong 0.453$.

\begin{figure}
\includegraphics[width =2.8in]{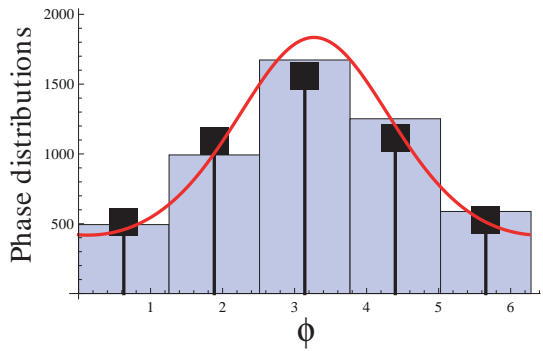}
\caption{Phase distribution for the same parameters as in Fig.~\ref{fig01}. Histogram of the phases of the oscillators, obtained from direct numerical simulations of Eq.~(\ref{AmplitudeOsc}), using the same numerical data  as shown in Fig.\ref{fig01}, but now plotting it as a histogram with 5 bars. The continuous curve corresponds to the steady state of the nonlinear Fokker-Planck equation (\ref{NLFPE}) obtained numerically (the same as shown in Fig.\ref{fig01}). The squares correspond to coarse-graining with $M=5$. In order to compare the results, we have used the normalization $\sum_{j=0}^{M-1}P_j \Delta\phi= 2\pi N/M$, with $M = 5$ (same as the number of bars) and $N=5000$, the number of oscillators considered in the simulation of  Eq.~(\ref{AmplitudeOsc}).}
\label{fig02}
\end{figure}

Although we can arbitrarily manipulate the number of states by using the function $f(r)$, the limiting case $M=3$ exhibits anomalies in the bifurcation structure that deserve a separate analysis. In the next subsection we discuss the 3-state case in detail. For $M\geq 4$ the transition to synchronization is described by the normal form for the mean field order parameter (see Appendix A)
\begin{equation}
\dot{R} = \left(\alpha_M -\beta_M\left|R\right|^2\right)R\label{NormalForm-2},
\end{equation}
where
\begin{align}
\alpha_M =& \left(\frac{\sin\Delta\phi}{2\Delta\phi}\right)f(0)\left(K - K_c\right),
\nonumber\\
\beta_M =& \left(\frac{\sin\Delta\phi}{2\Delta\phi}\right)K_c \left[\frac{\tan\left(\Delta\phi/2\right)}{\tan\Delta\phi}f(0) - f^{\prime}(0)\right].
\label{forApB}
\end{align}
Clearly, as $\Delta \phi \to 0$ or equivalently $M\to\infty$, $\alpha_M\rightarrow \alpha$ and $\beta_M \rightarrow \beta$ as defined in Eq.~(\ref{alphabeta}).
Moreover, the bifurcation picture is quite similar to that of the continuous phase oscillator. For decreasing $f$-functions the bifurcation is supercritical, while increasing $f$-functions may induce subcriticality. Figure~\ref{fig02} displays a comparison between the direct numerical simulation of Eq.~(\ref{AmplitudeOsc}) presented as a five-bar histogram, the steady state solution of the Fokker-Planck equation (\ref{NLFPE}) obtained numerically (red line), and the direct numerical simulation of the continuous-time Markov chain model for $M=5$ (black squares).  The 5-state periodic Markov chain gives a surprisingly good picture of the dynamics displayed by the model Eq.~(\ref{AmplitudeOsc}).

\subsection{The three-state case}
\label{sec:threestate}

%\begin{widetext}
\begin{figure*}[ht]
  \centering
 \subfloat[K=1.2\label{figa}]{\includegraphics[width=1.7in]{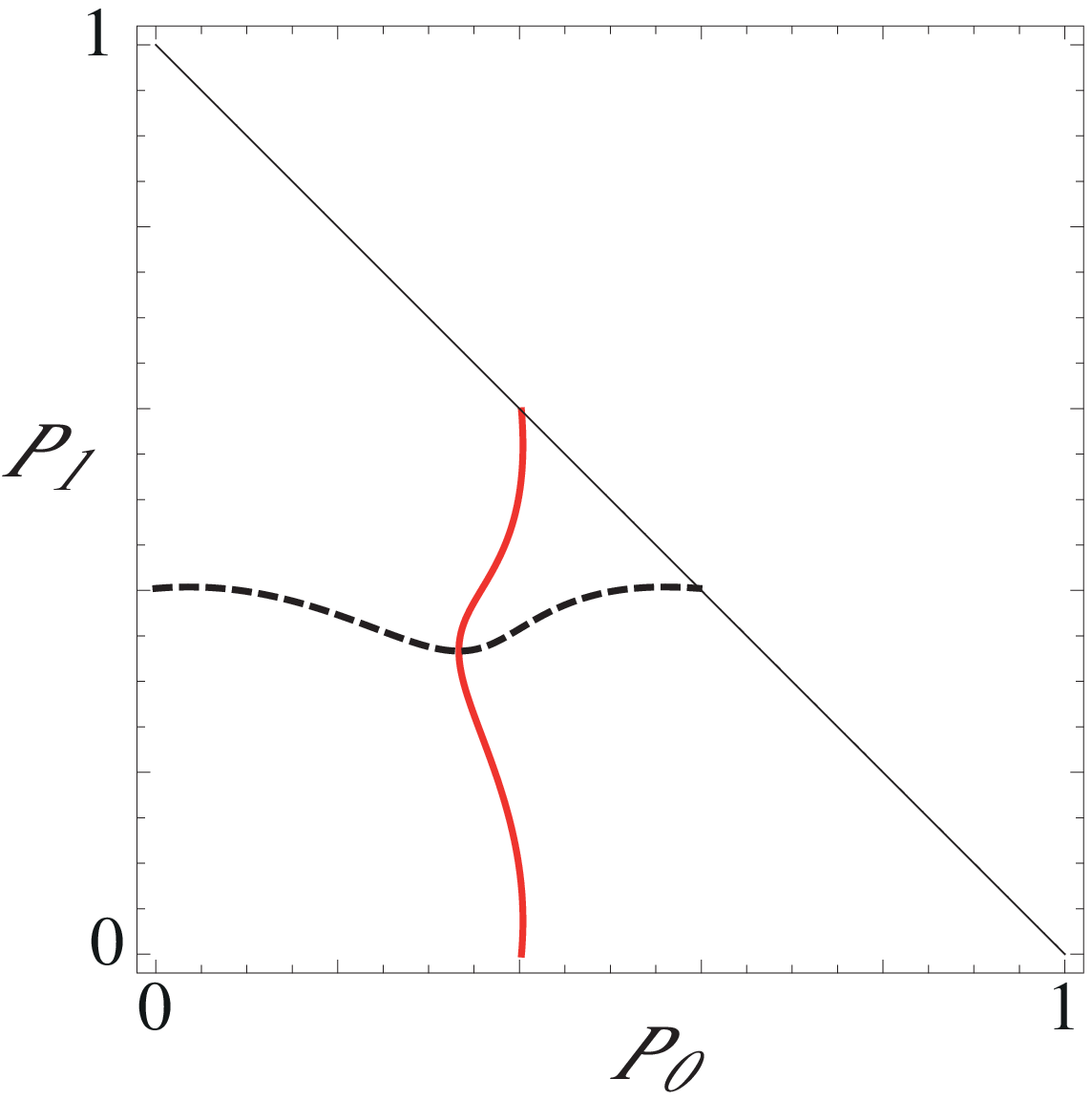}} %
 \subfloat[K=1.56\label{figb}]{\includegraphics[width=1.7in]{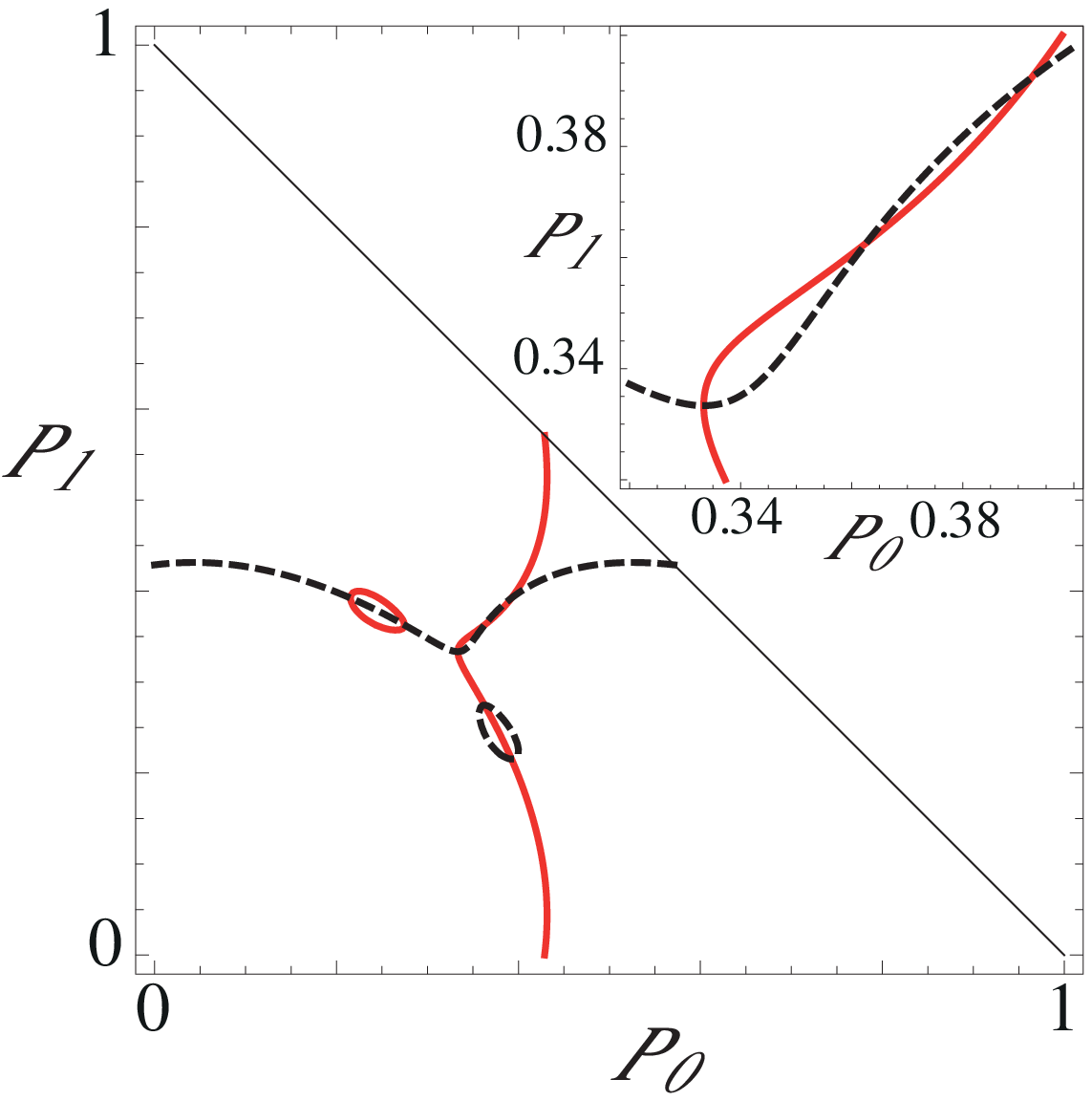}} %
 \subfloat[K=1.65399\label{figc}]{\includegraphics[width=1.7in]{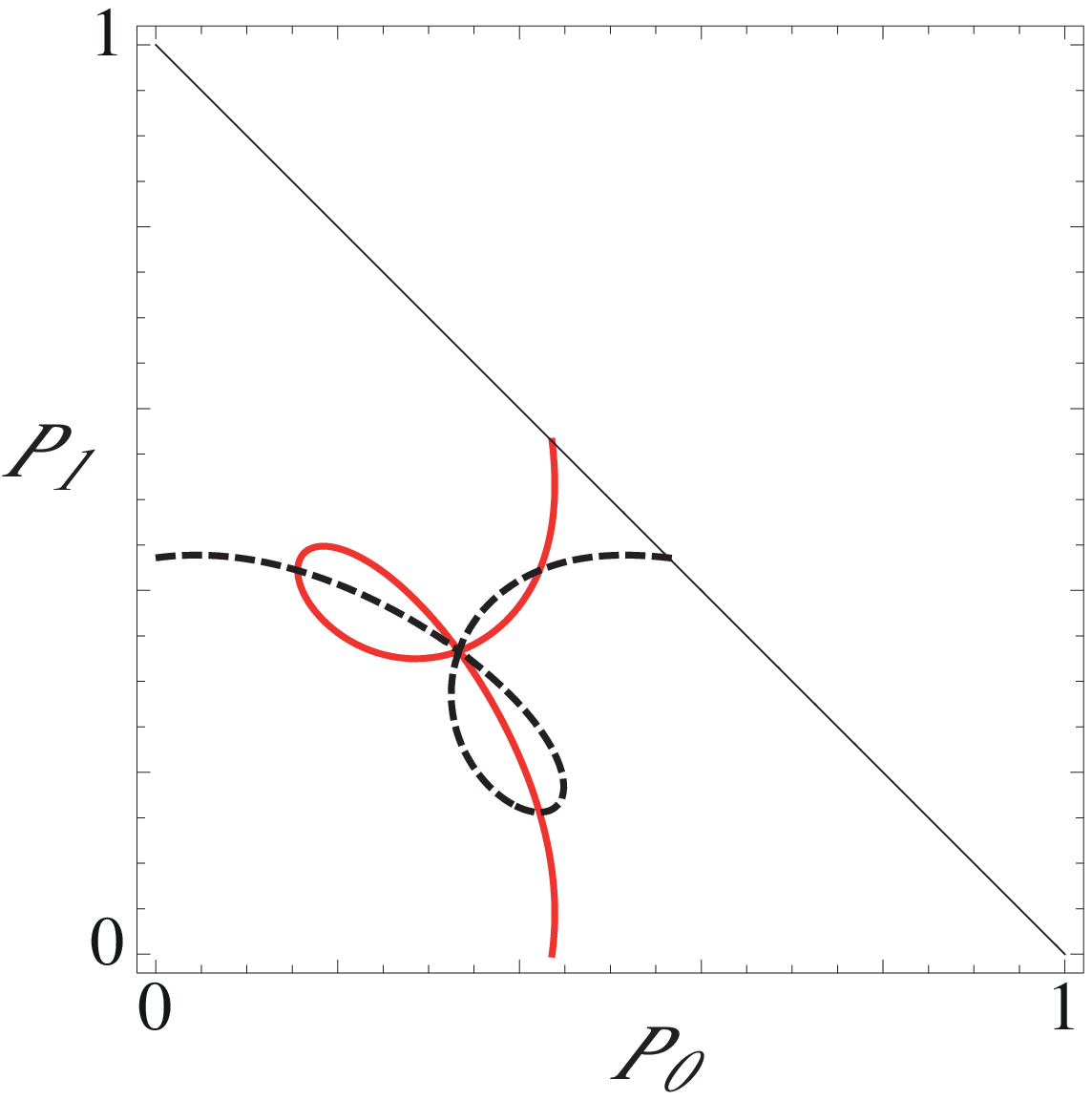}} %
 \subfloat[K=1.8\label{figd}]{\includegraphics[width=1.7in]{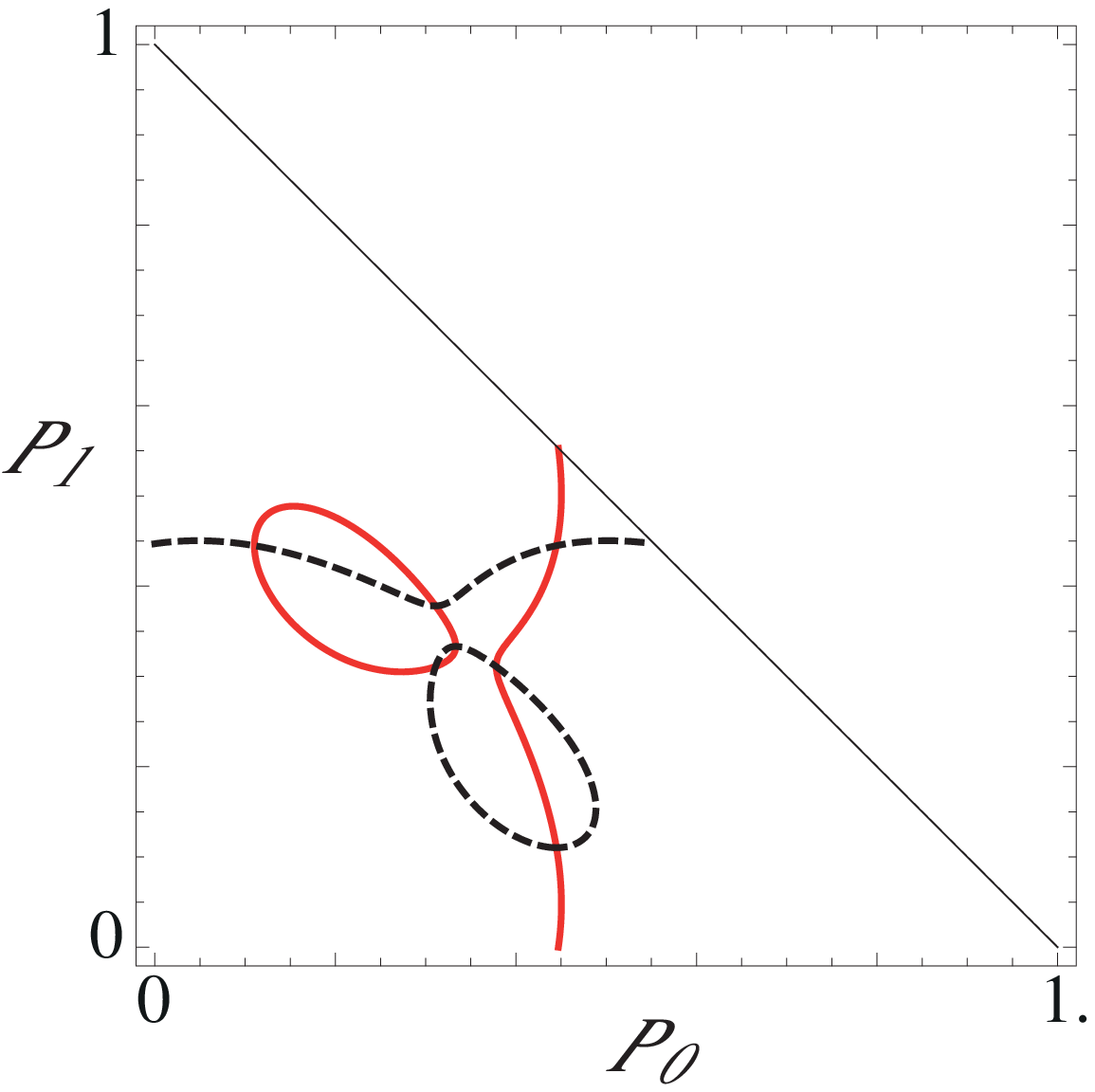}} %
  \caption{Nullclines of the nonlinear mean field master equation (\ref{NLME-2}), with $M=3$, $P_2 = 1 - P_0 - P_1$ and $\eta=1$, using the model function (\ref{fModel-1}) with $a=0.3$. The solid lines (red) show $G_0(P_0,P_1)=0$ in Eq.~(\ref{Gs}) and the dashed lines (black) show $G_1(P_0,P_1)=0$. From left to right: (a) $K<\widehat{K}_c$, a single stable fixed point, the desyncrhonized state, (b) $\widehat{K}_c<K<K_c$, seven fixed points, four of them stable, (c) $K=K_c$, four fixed points, one unstable and three stable,  and (d) $K >K_c$, three stable fixed points. The inset in the second panel clarifies the multiple crossings that contribute to the seven fixed points.}
\label{fig03}
\end{figure*}
%\end{widetext}

The case $M=3$ requires a separate treatment because its behavior is completely different from those of the discrete $M\geq 4$ oscillators. In the $M=3$  case the mean field order parameter near the critical point and at the lowest nonlinear order obeys the normal form (see Appendix A)
\begin{equation}
\dot{R} = \alpha_3 R - \gamma ({R^*})^2 
\label{NormalForm-3},
\end{equation}
where $R^*$ is the complex conjugate of $R$, and 
\begin{align}
 \alpha_3 =&\frac{3\sqrt{3} f(0)}{8\pi}\left(K - K_c\right),
\nonumber\\
 \gamma =& K_c f(0)\left(\frac{\sin\Delta\phi}{2\Delta\phi}\right)=\frac{27\eta}{8\pi^2},
\end{align}
which corresponds to a transcritical bifurcation.  Separating magnitude and phase, that is, $R=r e^{i\psi}$, we obtain 
\begin{align}
\dot{r} =&\alpha_3 r - \gamma r^2\cos\left(3\psi\right),
\nonumber\\
\dot{\psi} =& \gamma r\sin\left(3\psi\right).
\nonumber
\end{align}
The desynchronized state $r=0$ with an arbitrary phase $\psi$ is stable for $K<K_c$ and unstable for $K>K_c$. Also, here we have the unstable fixed points
\begin{align}
K < K_c ~~ \Rightarrow& ~~ r_{-}=\frac{-\alpha_3}{\gamma} ~~\text{and}~~ \psi_{-} =  \left\{\frac{\pi}{3}, \pi,\frac{5\pi}{3}\right\},
\nonumber\\
K > K_c ~~ \Rightarrow& ~~ r_{+}=\frac{\alpha_3}{\gamma} ~~~\text{and}~~ \psi_{+} =   \left\{0, \frac{2\pi}{3}, \frac{4\pi}{3}\right\}.
\nonumber
\end{align}
These are hyperbolic points in a two-dimensional phase space that undergo a phase shift from one of the angles in $\psi_{-}$ to one in $\psi_{+}$  when they cross the critical point (at the critical point $K=K_c$, $ r_{-}=r_{+}=0$ because $\alpha_3=0$). The unstable manifold of $\left\{r_{-}, \psi_{-}\right\}$ is unstable to perturbations of the modulus $r$, but stable to phase perturbations. In contrast, $\left\{r_{+}, \psi_{+}\right\}$ is stable to perturbations of the modulus $r$, but unstable to phase perturbations.

\begin{figure}
\includegraphics[width =2.7in]{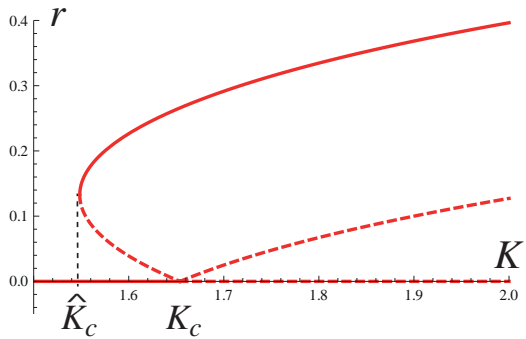}
\caption{Equilibrium $r$ as a function of $K$, for the same parameters as in Fig.\ref{fig03}. Continuous branches include all the stable fixed points, while dashed branches representing the unstable fixed points.}
\label{fig04}
\end{figure}

The nonlinear saturation of the instability occurs with the inclusion of higher nonlinear orders, which are not captured in the weakly nonlinear analysis used to deduce Eq.~(\ref{NormalForm-3}). We analyze it numerically and again use the model function (\ref{fModel-1}) for $f(r)$. Fig.~\ref{fig03} displays the nullclines of the dynamical system
\begin{align}
\dot{P}_0 =& G_0\left(P_0,P_1\right),
\nonumber\\
\dot{P}_1 =& G_1\left(P_0,P_1\right),
\label{Gs}
%\nonumber
\end{align}
where the nonlinear functions $G_0$ and $G_1$ are obtained from the mean field master equation (\ref{NLME-2}), with $M=3$ and the normalization condition $P_2 = 1 - P_0 - P_1$. The nullclines correspond to the curves $G_0\left(P_0,P_1\right) = 0$ and $G_1\left(P_0,P_1\right) = 0$ in the bidimensional phase space $\left(P_0,P_1\right)$. Moreover, due to the fact that $\left(P_0,P_1\right)$ are normalized probabilities, the physically accessible phase space is restricted to the right triangle defined by the region 
bounded by the lines $P_0 > 0 $, $P_1 > 0$, and  $P_0 + P_1 < 1$.
The fixed points correspond to the intersections of the two nullclines. Therefore, the bifurcation scenario is the following: For low enough coupling strength, the only stable equilibrium is the desynchronized state, $r=0$ (Fig.~\ref{figa}). At some point, say $K=\widehat{K}_c<K_c$, six new fixed points appear simultaneously by saddle-node bifurcation, three of them stable and three of them unstable. In the region $\widehat{K}_c < K < K_c$, there are four stable fixed points (Fig.~\ref{figb}). At the critical point $K=K_c$ (Fig.\ref{figc}), the three unstable fixed points collide with the desynchronized state, destabilizing it. For $K>K_c$, there remain only three stable fixed points, related to one another by synchronization with a phase shift $\Delta\phi$ (Fig.~\ref{figd}). The normal form (\ref{NormalForm-3}) only describes the collision between the desynchronized state and the three hyperbolic points.
A neater representation of the bifurcation scenario is shown in Fig.~\ref{fig04}, where we have plotted the equilibrium order parameter modulus of $r$ as a function of the coupling strength. The nonvanishing-$r$ branches represent three steady states with different phases $\psi$. For the parameters in this section, $\widehat{K}_c = 1.548\ldots$ and $K_c = 1.654\ldots$.

\subsection{Phase invariance in the discrete case}

Note that for $M=3$ (the three-state system) the phase invariance occurs in discrete steps $\Delta\phi$. This should be quite general and, from the structure of the expansion performed in Appendix A, one can see that, 
\begin{equation*}
\dot{\psi} \sim \Gamma \sin\left(M\psi\right),
\end{equation*}
where, in general, the coefficient $\Gamma$ should be computed for higher orders (perhaps thus revealing the existence of an extra time scale, slower than the one used for the expansion). In any case, it is intuitively rather clear that the distributions obtained from the discrete model will be invariant only to phase transitions occurring at discrete steps $\Delta\phi$.

\section{Summary and conclusions}
\label{sec:summary}

We have considered a noisy version of a Kuramoto-like model of identical continuous phase coupled oscillators that exhibit a transition to phase synchronization. We have addressed the following question: under what condition can we model this continuous dynamics as a discrete periodic continuous-time Markov chain, that is, as a discrete-phase model where the transitions between these discrete states are governed by a memoryless rate process? The states in the discrete chain represent a phase cluster of higher and higher density as the number of states in the chain decreases. 

The Markov chain model provides a surprisingly good description of the phase synchronization exhibited by the continuous model, even for chains of only five states (as confirmed with our numerical simulations), and we surmise that the same is true with at least four states. Reduction down to seven discrete states is possible with the interaction structure of the original Kuramoto model, that is, an interaction linear in the order parameter. Further reduction to six, five, four, and three states requires a generalization of the interaction to a nonlinear dependence on the order parameter.  However, we have shown that reduction to three-state exhibits more complex behavior in the bifurcation structure (that we analyzed in detail in Sec.~\ref{sec:threestate}) which are not present in the continuum model or in the discrete state models down to four states. 

Discrete stochastic models for synchronization phenomena have been increasing in popularity as a simple paradigm of synchronization \cite{Prager1,Prager2,Kouvaris1,Kouvaris2,Wood1,Wood2,Wood3,Wood4,Assis1,Assis2,Escaff1,Escaff2,Pinto,Escaff3,Rosas,Wood5}. Of course, this simplicity is related precisely to the relative ease of dealing with only a few states. In most of these discrete models there are only three states. Our results have shown that caution must be exercised when assuming too close a correspondence with Kuramoto-like continuous models and even with coarse-grained versions of the latter down to four states.

We are currently analyzing the behavior of continuous-time continuous-space oscillator arrays as well as Markov chains of four or five states for finite numbers of units. This introduces fluctuations that have not been considered in this work. It will be interesting to compare the effects of finite-size fluctuations in these various arrays. We are also considering arrays with negative coupling parameter $K$, a situation that we have analyzed for three-state Markov chains~\cite{Escaff3} and that leads to interesting dynamical structures.

\section*{Acknowledgements}

DE thanks FONDECYT Project No. 1140128 for financial support.  AR. acknowledges Capes for its support (Grant No.
99999.000296/2015-05). KL acknowledges the support of the U.S.
Office of Naval Research (ONR) under Grant No. N00014-13-1-0205. RT acknowledges FEDER (EU) and MINECO (Spain) 
under grant ESOTECOS FIS2015-63628-C2-R.

\bigskip

\appendix

\section{Critical point and normal form near criticality for continuous-time Markov chain model}

\bigskip

In this Appendix we detail steps that lead to three important equations in the coarse-graining process, namely, the critical coupling strength for synchronization [Eq.~(\ref{CriticalK-2})], the normal form of the evolution equation for the mean field order parameter that describes synchronization for $M\geq 4$ [Eq.~(\ref{NormalForm-2})], and the corresponding result for $M=3$ [Eq.~(\ref{NormalForm-3})].

\subsection{Critical Coupling}

To arrive at Eq.~(\ref{CriticalK-2}), we begin by analyzing the dynamics defined by Eq.~(\ref{NLME-1}), which is of the form
\begin{equation}
\dot{P}_j =\tilde{\eta} \mathcal{D}_2 P_j -  \frac{\tilde{K}}{f(0)}\mathcal{D}_1\left\{\Omega_j P_j\right\}
\label{eq:A1}
\end{equation}
where we define the operators $\mathcal{D}_1$ and $\mathcal{D}_2$ acting on any vector $\mathcal{F}$ with components $\mathcal{F}_j, j=0,1,\ldots,M-1$ and periodic boundary conditions ($\mathcal{F}_{0-1}=\mathcal{F}_{M-1}$ and $\mathcal{F}_{M-1+1}=\mathcal{F}_0$) as
\begin{align}
\mathcal{D}_1 \mathcal{F}_j &= \mathcal{F}_{j+1} - \mathcal{F}_{j-1},
\nonumber\\
\mathcal{D}_2 \mathcal{F}_j &= \mathcal{F}_{j+1} + \mathcal{F}_{j-1} - 2\mathcal{F}_j.
\nonumber
\end{align}
The coefficients in Eq.~(\ref{eq:A1}) are
\begin{equation}
\tilde{\eta} = \frac{\eta}{2(\Delta\phi)^2} ~~~\text{and}~~~ \tilde{K} = \frac{K f(0)}{2\Delta\phi}.
\label{Tilde}
\end{equation}

The array of oscillators is desynchronized and every state is equally likely, $P_j=1/M$, if the coupling $K$ is sufficiently weak. To find the critical coupling for synchronization we consider a perturbation of this state, 
\begin{equation*}
 P_j =  \frac{1}{M}\left[1 + \delta P_j \left(t\right)\right],
\end{equation*}
with 
\begin{equation*}
|\delta P_j |  \ll 1 ~~~\text{and}~~~ \sum_{j=0}^{M-1} \delta P_j = 0.
\end{equation*}
The inequality requires the perturbation to be small, and the sum ensures that the normalization of the $P_j$ is preserved.
We introduce this perturbation into Eq.~(\ref{eq:A1}) and retain terms up to first order in the perturbation:
\begin{equation}
\delta\dot{P_j} = \tilde{\eta} \mathcal{D}_2 \delta P_j -\tilde{K} \mathcal{D}_1 \left(\frac{1}{M} \delta\Omega_j +\left.\Omega_j\right|_{P_j=\frac{1}{M}}\delta P_j\right) .
\label{eq:Omega1}
\end{equation}
Here $\left.\Omega_j\right|_{P_j=\frac{1}{M}} = \Omega_j\left(R=0\right)$.
From Eqs.~(\ref{bound}) and (\ref{OPmf}),
\begin{equation*}
\delta \Omega_j = f(0) Im \left( \delta R e^{-i j \Delta \phi}\right)
\end{equation*}
and
\begin{equation*}
\delta R = \sum_{j=0}^{M-1} \delta P_j e^{ij\Delta \phi}.
\end{equation*}

We define the operator $\mathcal {L}$ as follows:
\begin{equation*}
\mathcal{L} F_j = Im\left(e^{-ij\Delta\phi}\frac{1}{M} \sum_{j^{\prime}=0}^{M-1}F_{j^\prime} e^{i j^{\prime} \Delta\phi} \right).
\end{equation*}
We can then write the evolution equation for $\delta P_j$:
\begin{equation}
\delta \dot{P_j} = \mathcal{A}~\delta P_j,
\label{Linear1}
\end{equation}
where the linear operator $ \mathcal{A}$ is given by
\begin{equation}
\mathcal{A} = \tilde{\eta}\mathcal{D}_2 - \tilde{K}\mathcal{D}_1\mathcal{L}.
\label{OLinear1}
\end{equation}

We next introduce the discrete orthogonal Fourier basis
\begin{equation}
\Psi_{mj} \left[C_m\right] \equiv C_m e^{-ijm\Delta\phi} +C^{*}_m e^{ijm\Delta\phi}.
\label{Fourier}
\end{equation}
The operator (\ref{OLinear1}) is Hermitian and diagonal in this basis under the inner product between two arbitrary $M$-element vectors, say $\left\{F_j\right\}_{j=0}^{M-1}$ and $\left\{G_j\right\}_{j=0}^{M-1}$:
\begin{equation}
\left\langle F_j\right.\left|G_j\right\rangle = \frac{1}{M} \sum_{j^{\prime}=0}^{M-1}F_jG_j .
\label{IntProduc}
\end{equation}
The associated eigenvalues are
\begin{equation}
\lambda_m  = -2\tilde{\eta} \left[1 - \cos \left(m\Delta\phi\right)\right] + \delta_{1m}\tilde{K}\sin\Delta\phi.
\label{eigenvalues}
\end{equation}
The eigenvalues for $m\geq 2$ are all negative.  The eigenvalue $\lambda_0=0$ is associated with the conservation of probability and has no dynamical consequences. Synchronization requires that at least one of the eigenvalues be positive. Only $\lambda_1$ can be positive, so $m=1$ is the critical mode and the critical point occurrs when $\lambda_1=0$, that is, when 
\begin{equation}
-2\tilde{\eta} \left[1 - \cos \left(\Delta\phi\right)\right] +\tilde{K}_c\sin\Delta\phi = 0.
\label{eigenvaluem1}
\end{equation}
Using Eq.~(\ref{Tilde}), this gives the critical coupling parameter
given in Eq.~(\ref{CriticalK-2}).
Therefore, the dynamics near criticality evolve on two time scales, a fast one associated with the relaxation of the modes with $m\geq 2$, and a slow one related with the critical mode $m=1$.

\subsection{Normal Form of the Order Parameter for $M\geq 4$}

Our next step is to derive the normal form of the evolution equation for the mean field order parameter that describes synchronization for $M\geq 4$ [Eq.~(\ref{NormalForm-2})].  
The order parameter in Eq.~(\ref{OPmf}) can be written as $R=M\left\langle P_j\right.\left|e^{i j \Delta\phi}\right\rangle$. Then, expanding $P_j$ in the basis (\ref{Fourier}), that is
\begin{equation*}
P_j = \sum_{m} \Psi_{mj} \left[C_m\right],
\end{equation*}
we find that 
\begin{equation*}
R=M\left\langle P_j\right.\left|e^{i j \Delta\phi}\right\rangle = MC_1.
\end{equation*}
Therefore, the order parameter $R$ is associated with the amplitude of the critical mode. That is, $R(t)$ is related to the slow time scale (central manifold).

We next introduce the ansatz 
\begin{equation}
 P_j = \frac{1}{M}\left(1 +  \sum_{n=1}^{\infty} W_{j}^{\left[n\right]} \left[R(t)\right]\right). 
 \label{NLexpansion}
\end{equation}
where the $W_{j}^{\left[n\right]}$ represents the contribution of order $n$ in $R$ and $R^{*}$. Note that, the conservation of probability implies that, for all $n$,
\begin{equation}
\langle W_{j}^{\left[n\right]}\left|\Psi_{0j}  \right\rangle = 0.
\label{ConserProb}
\end{equation}
$P_j (t) =P_j (R(t))$ evolves on the slow time scale $\lambda_1^{-1}$ because we are assuming that the fast contributions have already relaxed. That is,
\begin{equation}
\left|\lambda_1\right| \sim \left|K - K_c\right|\ll 1,\label{ScalingTime}
\end{equation}
which is small near criticality.

The next step is to introduce the ansatz (\ref{NLexpansion}) into Eq.~(\ref{eq:A1}) and separate order by order.
The first order, $n=1$, leads to the equation
\begin{equation}
\mathcal{A}_c  W_{j}^{\left[1\right]} = 0,
\label{O1}
\end{equation}
where
\begin{equation*}
\mathcal{A}_c = \tilde{\eta}\mathcal{D}_2 - \tilde{K}_c\mathcal{D}_1\mathcal{L}
\end{equation*}
is the linear operator (\ref{OLinear1}) at the critical point (\ref{CriticalK-2}). The solution of Eq.~(\ref{O1}) is
\begin{equation}
W_{j}^{\left[1\right]} = \Psi_{1j}\left[R\right].\label{W1}
\end{equation}
The higher orders, $n\geq 2$, lead to an infinite hierarchy of inhomogeneous linear equations,
\begin{align}
\text{order: 2} ~~~ \mathcal{A}_c W_{j}^{\left[2\right]} &= V_{j}^{\left[2\right]}\left(W_{j}^{\left[1\right]}\right),
\label{O2}\\
\text{order: 3} ~~~ \mathcal{A}_c W_{j}^{\left[3\right]} &= V_{j}^{\left[3\right]}\left(W_{j}^{\left[1\right]},W_{j}^{\left[2\right]}\right),
\label{O3}\\
... & ...
\nonumber\\
\text{order: n} ~~~ \mathcal{A}_c W_{j}^{\left[n\right]} &= V_{j}^{\left[n\right]}\left(W_{j}^{\left[1\right]}, ..., W_{j}^{\left[n-1\right]}\right).
\label{On}\\
...& ...
\nonumber
\end{align}
Therefore, near criticality we have transformed the nonlinear equation Eq.~(\ref{eq:A1}) into an infinite hierarchy of linear equations. The functions $V_{j}^{\left[n\right]}$ depend on the previous orders in $n$ and  should be computed order by order.
We must be careful with the fact that the operator $\mathcal{A}_c$ is not invertible since it has the nontrivial kernel $\mathcal{A}_c \Psi_{0j}=\mathcal{A}_c \Psi_{1j}=0$. Therefore, to ensure the consistency of the expansion (\ref{NLexpansion}), we must impose a solvability condition at all orders. Since $\mathcal{A}_c$ is Hermitian under the inner product (\ref{IntProduc}), we may use Fredholm's alternative~\cite{ELPHICK}, which leads to the solvability conditions
\begin{align}
\left\langle \Psi_{0j}\left[C_0\right] \right. | V_{j}^{\left[n\right]} \rangle &=  0,
\label{SC0}\\
\left\langle \Psi_{1j}\left[C_1\right] \right. | V_{j}^{\left[n\right]} \rangle &=  0.
\label{SC1}
\end{align}
The first solvability condition, Eq. (\ref{SC0}), is directly implied by the conservation of probability. It is therefore trivially satisfied at all orders. In contrast, the solvability condition (\ref{SC1}), which is related to the critical mode, has nontrivial implications. We will use Eq. (\ref{SC1}) to compute the normal forms.

For the order parameter we assume the pitchfork bifurcation scaling
\begin{equation}
 \left|R\right| \sim \left|K - K_c\right|^{1/2}  \ll 1
 \label{ScalingPitch}
\end{equation}
and check the consistency of this assumption \emph{a posteriori}.  
Assuming (\ref{ScalingPitch}) and (\ref{W1}), the second order, $n=2$, leads to 
\begin{equation}
 \mathcal{A}_c W_{j}^{\left[2\right]}=V_{j}^{\left[2\right]} = -\tilde{K}_c \sin\left(2\Delta\phi\right)\Psi_{2j}\left[R^2\right].\label{V2}
\end{equation}
For $M>3$, Eq.~(\ref{V2}) does not have solvability problems, leading to 
\begin{equation*}
W_{j}^{\left[2\right]} =  \Psi_{2j}\left[\frac{2\tan\left(\Delta\phi/2\right)}{\tan\left(\Delta\phi\right)}R^2\right].
\end{equation*}
The third order, $n=3$, has solvability problems, and the solvability condition (\ref{SC1}) leads to the equation
\begin{align}
  & -2\dot{R} +2\Delta\tilde{K} \sin\left(\Delta\phi\right)R 
\nonumber\\
~~~&- 2\tilde{K}_c \sin\left(\Delta\phi\right)\left[ \frac{\tan\left(\Delta\phi/2\right)}{\tan\left(\Delta\phi\right)} -\frac{f^{\prime}(0)}{f(0)}\right]\left|R\right|^2 R &= 0
\nonumber.
\end{align}
Using Eq.~(\ref{Tilde}), the above solvability condition takes the form given in Eq.~(\ref{NormalForm-2}).

\subsection{Normal Form of the Order Parameter for $M=3$}

For $M=3$, Eq.~(\ref{V2}) has no solution, since in this case $\Delta\phi=2\pi/3$, which implies
\begin{equation*}
\Psi_{2j}\left[R^2\right] = \Psi_{1j}\left[(R^{*})^2\right].
\end{equation*}
Here the scaling assumption (\ref{ScalingPitch}) does not allow us to impose a suitable solvability condition at second order. Hence, in order to ensure the consistency of the expansion (\ref{NLexpansion}), we must modify our scaling assumption and adopt the transcritical scaling 
\begin{equation}
\left|R\right| \sim \left|K - K_c\right|  \ll 1.\label{TransScaling}
\end{equation}
This scaling allows us to write the solvability condition (\ref{SC1}) for Eq.~(\ref{O2}) in the form
\begin{equation*}
  -2\dot{R} +2\Delta\tilde{K} \sin\left(\Delta\phi\right)R - 2\tilde{K}_c \sin\left(\Delta\phi\right)({R^*})^2 = 0.
\end{equation*}
Using Eq.~(\ref{Tilde}), the above solvability condition leads to the normal form Eq.~(\ref{NormalForm-3}).

\medskip

\section{Brief commentaries about the continuos phase case.}

\medskip

Critical point calculations and normal forms near the transition to synchronization for the continuos-phase Kuramoto model and its variants have been extensively documented in the literature~\cite{Acebron,Kuramoto2,Bonilla}  (See Ref.~\cite{Acebron} for an extensive review of a number of approaches). 
%As well as alternative method, as, for example, the Chapman-Enskog expansion of kinetic theory applied to an inertial variation of the nonlinear Fokker-Planck equation (\ref{NLFPE}), instead the Boltzmann transport equation \cite{Bonilla}. 
Here we simply point out that in the limit $\Delta\phi\rightarrow 0$  some of the results of Appendix A reduce to those appropriate for the continuos-phase oscillators. In particular, in this limit,
\begin{equation}
\lim_{\Delta\phi\rightarrow 0}K_c = \frac{\eta}{f(0)},
\label{CriticalK-Apen2}
\end{equation}
that is, we obtain the continuous critical point (\ref{CriticalK-1}). Moreover, from Eq.~(\ref{forApB}) we find
\begin{align}
\lim_{\Delta\phi\rightarrow 0}\alpha_M =\alpha =& \frac{f(0)}{2}\left(K - K_c\right),
\nonumber\\
\lim_{\Delta\phi\rightarrow 0}\beta_M =\beta =&  \frac{K_c}{2} \left[\frac{1}{2}f(0) - f^{\prime}(0)\right],
\nonumber
\end{align}
which leads to the normal form given in Eq.~(\ref{NormalForm-1}).
When comparing our results to those reported in the literature, in addition to the  limit $\Delta\phi\rightarrow 0$  we stress that here we are working with identical oscillators [$g(\omega)=\delta(\omega)$] and that in the literature on the Kuramoto model the function $f(r)=1$.

To obtain the analytic estimate (\ref{NFapprox}) of the steady state distribution, we note that for small $\Delta\phi$, 
\begin{equation*}
P_j(t) \approx \rho(j\Delta\phi,t)\Delta\phi.
\end{equation*}
Then, retaining the lowest order of the expansion (\ref{NLexpansion}),
\begin{equation*}
 \rho(j\Delta\phi,t) \approx \frac{1}{2\pi}\left(1 +  W_{j}^{\left[1\right]}\right),
\end{equation*}
Next, we take the continuos limit $\Delta\phi\rightarrow 0$ or $M\rightarrow\infty$, which implies $j\Delta\phi\rightarrow\phi$, with the solution (\ref{W1}). The function  $\Psi_{1j}\left[R\right]$ is obtained from the definition (\ref{Fourier}). Therefore, at the steady state,
\begin{equation*}
 \rho_{st}(\phi) \approx \frac{1}{2\pi}\left(1 +  2 Re\left[R_{st}e^{-i\phi} \right] \right),
\end{equation*}
where the steady state value $R_{st}$ of the order parameter is estimated from the equilibrium value predicted by the normal form  (\ref{NormalForm-1}) for $K>K_c$.  That is,
\begin{equation*}
r_{st} = \sqrt{\frac{\alpha}{\beta}} ~~~\text{and hence}~~~ R_{st} = \sqrt{\frac{\alpha}{\beta}}e^{i\psi},
\end{equation*}
where $\psi$ is an arbitrary phase constant. Using the model function (\ref{fModel-1}) for $f(r)$, we find that
\begin{equation*}
r_{st} = \sqrt{\frac{2a(K-\eta)}{\eta(a+2)}}.
\end{equation*}
Therefore we obtain Eq. (\ref{NFapprox}).

\end{document}